# Hamiltonian SU(N) Lattice Gauge Theories With Exact Vacuum States In (2+1) Dimensions

George M. Frichter

*Kansas Institute for Theoretical and Computational Science, University of Kansas,*

*Lawrence, KS 66045*

D. Robson

*Department of Physics, Florida State University, Tallahassee, FL 32306*

(December 9, 1993)

## Abstract

We investigate (2+1)-d Hamiltonian lattice gauge theory using a class of Hamiltonians having exactly known vacuum states. These theories are shown to have a wide range of possible classical continuum limits which differ from that of the standard Kogut-Susskind Hamiltonian. This conclusion is at variance with some previously published results. We examine the quantum continuum behavior of these theories by both analytic and numerical methods including plaquette space integration and standard Monte Carlo techniques. String tension and variational estimates for the $J^{\text{PC}} = 0^{++}$ glueball spectra are presented for SU(3). We find that in spite of the wide range of classical behavior predicted, these theories correspond to only two distinct quantum systems in the weak coupling limit. One of these quantum limits gives string tensions and glueball states which show scaling in weak coupling which agrees with the perturbative prediction for the (2+1)-d problem.

Typeset using REVTeX

# I. INTRODUCTION

Our current understanding of the strong force gives us a picture of fermion fields interacting via gauge fields generated by an underlying local SU(3)$_{\text{color}}$ symmetry. Since the physical hadrons are observed to be color singlets, this picture requires exact color confinement. We also know that strong interaction processes involving large momentum transfer are characterized by nearly free hadronic constituents. The current belief is that non-Abelian gauge theories such as SU(3)$_{\text{color}}$ simultaneously possess the features of asymptotic freedom for short distance phenomena and exact color confinement. Clarification of the low energy behavior of SU(3)$_{\text{color}}$ gauge theory requires a nonperturbative approach and is surely one of the most fundamental problems in the field of strong interaction physics.

Truly accurate lattice calculations of glueball masses without dynamical fermions is an essential first step toward the long range goal of describing completely the hadronic spectrum from QCD. Wilson's [1] nonperturbative formulation of gauge theory on a space-time lattice was developed as a means of controlling ultraviolet divergences inherent in QCD. Truncation of the lattice then yields a theory having a finite number of degrees of freedom making direct computer simulation of the system feasible. Soon afterward, an alternative Hamiltonian formulation of the problem was given by Kogut and Susskind [2]. In this canonical form of the theory, physical observables such as glueball masses and string tension appear directly as eigenvalues, and in addition, the gluon degrees of freedom reside on the links of a three dimensional space lattice instead of the four dimensional Euclidean space-time lattice used in Wilson's approach.

A potentially important feature of the lattice Hamiltonian is that it may not be unique. The common wisdom is that it must be locally gauge invariant and possess a classical continuum limit in accord with the continuum gauge theory it approximates. Beyond these rather general constraints, the precise form of the Hamiltonian is not determined, and this freedom can be used to fix the vacuum wavefunction, resulting in a theory having a Hamiltonian modified relative to the standard Kogut-Susskind (KS) form [3] [4]. This is the approach we have adopted here. It represents a potentially important advance in the Hamiltonian formulation of lattice gauge theory since it avoids the inherent computational complexity and limited numerical accuracy of either dynamically evolving the exact vacuum state using Monte Carlo techniques [5], or introducing into the formalism an approximate variational ansatz for the vacuum [5] [6] [7] [8] [9]. With an exact vacuum in hand, variational estimates of the glueball mass spectra become rigorous upper bounds because the vacuum energy is precisely known.

In this paper, we investigate a class of (2+1)-dimensional lattice Hamiltonians for the SU(N) gauge theory which have an independent-plaquette vacuum structure. Achieving a theory wherein the vacuum is exact involves adding a term, $\Delta H$, to the KS Hamiltonian. The notation here is suggestive of the hope that this modification is small in some sense and leads to a theory consistent with the corresponding continuum gauge theory. The fact that the exact vacuum has a simple structure with no long range magnetic correlations has been the source of concern to some authors [7] [10], however, we believe that equally compelling arguments can be made in favor of the present formalism [4].

We view the classical limit of our lattice Hamiltonian as an indicator that perhaps the present theory can agree with continuum QCD. In the appendix, we demonstrate that $\Delta H$ remains finite in the classical continuum limit of the (2+1)-d theory. One also finds that this term has a vanishing limit in the more physically relevant (3+1)-d version of the theory. This is encouraging since, after all, we think of the KS Hamiltonian as a valid discrete representation of the continuum theory based largely on this same classical correspondence. We keep in mind, however, that we are ultimately interested in clarifying the quantum behavior of these systems particularly in the scaling region.

Also in support of the present approach, consider the key role played by the independent-plaquette component of the vacuum in the KS theory. Several variational calculations have been reported which clearly support the notion of a vacuum dominated by long range magnetic disorder even well into the scaling region [5] [6] [7] [8] [9]. While this simple picture can not be correct in the extreme weak coupling limit, it does seem to capture the main features of the true KS vacuum in regions where scaling is observed and is thought to be everywhere exact as a long wavelength representation [11] [12] [13]. Again, this allows one to argue that at least in the near-weak coupling regime, $\Delta H$ represents a small adjustment of the KS dynamics which ensures an independent-plaquette vacuum.

In this paper, we present numerical results for the SU(3) version of the theory using both exact (expectation values of gauge invariant operators are exactly integrable in the (2+1)-d theory due to the simple Jacobian) and standard Monte Carlo methods. We make use of Monte Carlo mehtods since we are ultimately interested in the (3+1)-d version of the SU(3) theory (where a complicated Jacobian involving Bianchi identities appears) and wish to calibrate our computer codes where exact results are possible.

To further motivate the present approach, we have also obtained the SU(2) string tension using our formalism which we will now compare with previously published results obtained by more conventional means. Teper [14] [15] has recently obtained $\beta\sqrt{\sigma} = 1.336 \pm 0.01$ (see reference for notation) for SU(2) string tension within the $D = 3$ Euclidean theory. This compares very well with our result of $1.30 \pm 0.1$ (approximate error estimate) for this same quantity. Using a one and two-plaquette varational ansatz for the KS vacuum, Arisue $et\ al$ [6] obtained $\beta\sqrt{\sigma} = 1.61$ for the independent-plaquette vacuum and 1.38 for the two-plaquette vacuum. These values are eyeball estimates from their figure 5 in the scaling region near $\beta = 1.86$. Comparision of our result with these variational results clearly suggests that our exact local vacuum formalism is doing more than a one-plaquette ansatz does within the KS theory. We are very encouraged by this result and proceed to further assess the strengths and weaknesses of this potentially important approach.

The paper is organized as follows. In section II and the appendix, we discuss the classical continuum limits of a family of lattice Hamiltonians having exactly known vacuum states. Section III discusses string tension for the present theory by examining exact results for the plaquette expectation value. In section IV, we describe the variational approach to the $0^{++}$ glueball spectra and in section V we give the specifics of the Monte Carlo simulations. An examination of our numerical results appears in section VI and finally, section VII summarizes the main points and discuss prospects for future investigation.

## II. LATTICE HAMILTONIANS WITH EXACT VACUUM STATES

The Kogut-Susskind [2] form of the lattice Hamiltonian for SU(N) gauge theory can be derived in a straightforward way from Wilson's lattice action using the standard canonical prescription for obtaining a Hamiltonian from a Lagrangian [16]. The result is,

$$H_{\text{KS}} = \frac{g^2}{2a} E_l^\alpha E_l^\alpha + \frac{1}{ag^2} \sum_p [2N - \text{Tr}(U_p + U_p^\dagger)] . \tag{1}$$

Here, $g$ is the lattice coupling constant, $a$ is the lattice spacing, $l$ labels the links upon which gauge degrees of freedom (SU(N) rotations) reside and serve to join neighboring lattice sites, $\alpha$ is a color label, $E_l^\alpha$ are color-electric field operators representing the variables conguate to the gauge (link) variables, $p$ labels the plaquettes, and finally, $U_p$ are plaquette variables. A plaquette is defined as the path-ordered product of link variables, $U_l$, obtained by traversing a closed square loop in the lattice having one link to a side. Sums over repeated indices are assumed throughout this paper.

Being conjugate variables, the link degrees of freedom and the color-electric field operators satisfy equal time commutation relations with respect to each other, i.e.,

$$[E_l^\alpha, U_{l'}] = \delta_{ll'} \frac{\lambda^\alpha}{2} U_l \text{ and } [E_l^\alpha, U_{l'}^\dagger] = -\delta_{ll'} U_l^\dagger \frac{\lambda^\alpha}{2} . \tag{2}$$

We can see that $E_l^\alpha$ acts as a differential operator on the variable $U_l$ with respect to the parameters of the SU(N) rotation. It follows then that the first term of Eq. (1) is a sum of the kinetic energies possessed by the link variables, and the second is a sum over potential terms which depend on the relative orientation of the gauge degrees of freedom through couplings provided by the plaquettes.

In the limit of small lattice spacing, $H_{\text{KS}}$ agrees with the classical Hamiltonian for the continuum SU(N) gauge theory. In 3 space dimensions the result is very familiar,

$$H_{\text{KS}} \xrightarrow{a \to 0} \frac{1}{2} \int (\vec{E}^\alpha \cdot \vec{E}^\alpha + \vec{B}^\alpha \cdot \vec{B}^\alpha) d^3x . \tag{3}$$

Agreement of the classical limit is usually regarded as a reasonable minimum condition that a lattice approximation should satisfy. However, we are ultimately interested in whether the quantum dynamics generated by various lattice approximations such as Eq. (1) can yield the known results from the continuum theory. Thus, we look to the scaling behavior of the string tension and glueball masses emerging in the weak coupling limit of our lattice calculations for a more meaningful indication of whether a particular lattice formulation is capable of producing physical results or not.

To what extent then is the classical continuum limit indicative of the quantum continuum behavior? This is the question of universality in lattice gauge theory. We will demonstrate that a particular class of lattice Hamiltonian whose members have a wide variety of possible classical limits, correspond to quantum systems which have only two distinct weak coupling limits. One of these weak coupling limits appears to be the physical one and is in agreement with perturbative predictions from the continuum gauge theory.

The Hamiltonians we will examine have the desirable feature that their vacuum states are exact for all values of the lattice coupling, thereby making the calculation of vacuum expectation values of various lattice operators, which are needed for the evaluation of string tension and variational estimates of glueball masses, a relatively easy task. An exact vacuum can be achieved by adding a term, $\Delta H$, to the basic KS Hamiltonian,

$$H = H_{\text{KS}} + \Delta H \ . \tag{4}$$

At this point, it is convenient to redefine the energy scale by subtracting the constant, ultraviolet divergent part of the magnetic term in Eq. (1). Also, in (2+1) dimensions, one finds that the lattice coupling is related to the dimensionless gauge coupling, $e$, by $g^2 = ae^2$. We then have,

$$H = \frac{e^2}{2} E_l^\alpha E_l^\alpha - \frac{1}{a^2 e^2} \sum_p \text{Tr}(U_p + U_p^\dagger) + \Delta H \ . \tag{5}$$

We will choose $\Delta H$ such that the modified Hamiltonian is gauge invariant, Hermitian, and positive definite.

We will now look for a form for $\Delta H$ which renders a particular vacuum state exact. Consider a vacuum of the form,

$$|\Psi_0\rangle = e^R|0\rangle, \qquad (6)$$

where $|0\rangle$ is the product of individual link ground states defined by $E_l^\alpha|0\rangle = 0$. In this paper, the form we consider for $R$ is,

$$R = \sum_p (\beta_1 A_p + \beta_2 A_p^2), \qquad (7)$$

with

$$A_p = \text{Tr}(U_p + U_p^\dagger). \qquad (8)$$

Here, $\beta_1$ and $\beta_2$ are as yet undetermined functions of the lattice coupling. If $H$ is positive definite and $H|\Psi_0\rangle = 0|\Psi_0\rangle$, then $|\Psi_0\rangle$ is assured of being the vacuum. Using this condition to determine $\Delta H$, one finds,

$$\Delta H|\Psi_0\rangle = [-\frac{e^2}{2}(E_l^\alpha E_l^\alpha R) - \frac{e^2}{2}(E_l^\alpha R)(E_l^\alpha R) + \frac{1}{a^2 e^2}\sum_p A_p]|\Psi_0\rangle. \qquad (9)$$

Thus, $H$ has the same classical continuum limit as the KS theory if,

$$\Delta H = -\frac{e^2}{2}(E_l^\alpha E_l^\alpha R) - \frac{e^2}{2}(E_l^\alpha R)(E_l^\alpha R) + \frac{1}{a^2 e^2}\sum_p A_p \xrightarrow{a \to 0} 0. \qquad (10)$$

We view this condition as a guide in searching for theories which yield the same physics as the standard KS form, keeping in mind that we are really interested in the quantum behavior of the resulting systems. In fact, one of the results of the present work is that theories need not share classical limits to give identical physical results. In the appendix to this paper, we look for $\beta_1$ and $\beta_2$ such that Eq. (10) is true. We find that it can not be satisfied for the (2+1)-d problem but that $\Delta H$ can at least be made finite by imposing the constraint,

$$\beta_1 + 4N\beta_2 = \frac{1}{2C_N g^4}, \qquad (11)$$

along with the condition that $\beta_2$ diverges no faster than $a^{-2}$,

$$\beta_2 \leq O(a^{-2}) \,. \tag{12}$$

These conditions eliminate divergences in $\Delta H$ as the lattice spacing vanishes. However, as we discuss in that section, they are not enough to fix the classical limit of $\Delta H$. One must further specify how $\beta_1$ and $\beta_2$ individually behave as the lattice coupling and hence the lattice spacing are taken to zero. If we relax the latter condition and only enforce Eq. (11), then we have shown in the appendix that the resulting Hamiltonians may differ from the KS theory by a divergent amount.

In the next section, we will numerically examine some theories which satisfy Eqs. (11) and (12). Specifically, we put $\beta_1 = \frac{\mathcal{X}}{2C_N g^4} + b$ and $\beta_2 = \frac{1}{4N}[\frac{1-\mathcal{X}}{2C_N g^4} - b]$ and consider the continuum limits for fixed values of $\mathcal{X}$ and $b$. Figure 1 indicates the relationships among the variables we have introduced thus far for the case $b = 0$. In the appendix, we find the classical limit of our modified Hamiltonian to be independent of $b$,

$$H \xrightarrow{a \to 0} \frac{1}{2} \int (\vec{E}^\alpha \cdot \vec{E}^\alpha + \frac{1+3\mathcal{X}}{4}\vec{B}^\alpha \cdot \vec{B}^\alpha)d^2x + \frac{1}{8C_N^2 e^4} \int d^2x [(\nabla \times \vec{B}^\alpha) \cdot (\nabla \times \vec{B}^\alpha)] \,. \tag{13}$$

So, the magnetic component of the resulting Hamiltonian is modified relative to the KS form by an amount which depends on the relative strengths of the $\beta_1$ and $\beta_2$ terms through the parameter $\mathcal{X}$. In particular, the critical value $\mathcal{X}_c = -\frac{1}{3}$ marks a sign change in the magnetic contribution to the Hamiltonian. We also see a contribution which is independent of the $\beta$'s so long as Eq. (11) is satisfied.

In spite of the widely diverse classical continuum behaviors for these theories, we will now argue (and later demonstrate numerically) that Eq. (11) alone is sufficient to ensure only two possible weak coupling limits for the corresponding quantum lattice systems.

Consider the continuum limit of the vacuum state density $\Psi_o^2 = e^{2R}$. This limit will determine the continuum behavior of observables such as the string tension as the lattice cutoff is removed. From Eqs. (7) and (11), we can see that as the lattice coupling is taken to zero, the extrema of $R$ will dominate integrals over the plaquette degrees of freedom (stationary phase condition). In the case of SU(3), there are three turning points for $R$ corresponding to $A_p = 6$, $A_p = -2$, and $A_p = -3$. Which one 'wins' in the limit of weak coupling simply depends on which density is greater, $\Psi_o^2(A_p = 6)$, $\Psi_o^2(A_p = -2)$, or $\Psi_o^2(A_p = -3)$. One can easily verify using Eq. (16) below that the extreme values, $A_p = 6$ or $A_p = -3$, always dominate the integration. Thus there will be only one phase change occuring at the crossover point for these two solutions,

$$R(A_p = 6) = R(A_p = -3)$$
$$\beta_1 + N\beta_2 = 0 \,, \tag{14}$$

or in terms of $\mathcal{X}$ and $b$,

$$\mathcal{X} + 2bC_N g^4 = -\frac{1}{3} \,, \tag{15}$$

where the critical value $\mathcal{X}_c = -\frac{1}{3}$ is given by the limit of weak coupling ($g \to 0$) and corresponds to the sign change in the magnetic part of the Hamiltonian seen in Eq. (13). We see then that $\mathcal{X}_c$ separates two distinct regions of quantum behavior. For example, when $\mathcal{X} < \mathcal{X}_c$, the vacuum expectation value $\langle A_p \rangle_o$ approaches $-3$ as the lattice approximation is removed whereas $\mathcal{X} > \mathcal{X}_c$ gives a limit in agreement with the classical value of $+6$ (see Eq. (A6)). In the next section, we present numerical results which are in complete agreement with the above analysis.

So, here is the situation. Each value of $\mathcal{X}$ gives a family of lattice Hamiltonians, corresponding to different values of $b$, which share a unique classical limit as well as a unique limit for the associated quantum system. This supports the concept of universality in the sense that all Hamiltonians we have studied which share the same classical limit also yield identical physics. However, the mapping between classical and quantum limits is not one-to-one. Classical theories corresponding to different values of $\mathcal{X}$ map onto only two distinct quantum systems in the weak coupling limit. The quantum limit obtained in a particular case corresponds directly to the sign of the magnetic contribution to the Hamiltonian in the classical limit. Thus we see that many different classical Hamiltonians correspond to the same quantum system. The relative strength of the electric and magnetic parts of the classical Hamiltonian plays no part in the continuum limit of the quantum system; only the sign matters.

We now express Eq. (7) in terms of the parameters $\mathcal{X}$ and $b$ to be used in the subsequent sections. Also, for the remainder of this paper, we will work exclusively with SU(3) gauge theory. If we define the lattice coupling parameter, $\xi = \frac{1}{Ng^2}$,

$$R = \sum_p \left[\frac{27\xi^2}{8}(\mathcal{X} + \frac{1-\mathcal{X}}{12}A_p) + b(1 - \frac{1}{12}A_p)\right]A_p . \tag{16}$$

Before proceeding with the numerical results, we should point out differences between conclusions reached in this section and the appendix to this paper and those arrived at previously by other authors [3] [4] working with the same Hamiltonians and proposed vacuum states used here. The condition given by Eq. (11) is not the same as either of the two conditions suggested in the work of Ref. [3] except for the special case of $\beta_2=0$. In this special case their conditions on $\alpha_2$ (our $\beta_1$) given by their Eqs. 2.16 and 2.17 are inconsistent, i.e., $\alpha_2 = \frac{1}{(2C_N g^4)}$ and $\alpha_2 = 0$ respectively. We believe their constraints to be erroneous except for their special case of $R_1 = \alpha_1 \sum_p A_p$ with $\alpha_1 = \frac{1}{(2C_N g^4)}$. Indeed in this latter case we do reproduce their calculations for the mass gap (their Fig. 8) and string tension (their Fig. 9). The correct treatment of the classical continuum limit is presented in the appendix and is well verified numerically here in the results presented in the next section.

### III. SU(3) STRING TENSION

We are interested in evaluating the quantity,

$$\langle A_p \rangle_o = \frac{\int A_p \Psi_o^2 \{dU_l\}}{\int \Psi_o^2 \{dU_l\}} . \tag{17}$$

Since the Jacobian $\mathcal{J}$ for the change of variables $\{dU_l\} \to \mathcal{J}\{dU_p\}$ is unity for the (2+1)-d problem and the integrand is explicitly a function of $U_p$, the expectation value above reduces to an integral over a single plaquette variable,

$$\langle A_p \rangle_o = \frac{\int A_p e^{\sum_p [\frac{27\xi^2}{4}(\mathcal{X}+\frac{1-\mathcal{X}}{12}A_p)+2b(1-\frac{1}{12}A_p)]A_p} dU_p}{\int e^{\sum_p [\frac{27\xi^2}{4}(\mathcal{X}+\frac{1-\mathcal{X}}{12}A_p)+2b(1-\frac{1}{12}A_p)]A_p} dU_p} \tag{18}$$

where we have used Eq. (16). For SU(3), these two integrals can be reduced to two dimensions using $U_p$ in diagonal form and evaluated directly for various fixed values of $\mathcal{X}$, $b$, and $\xi$.

Figure 3 shows a portion of the surface $\langle A_p \rangle_o(\mathcal{X}, \xi)$ with $b = 0$ and clearly shows an approach to a discontinuity across $\mathcal{X}_c = -\frac{1}{3}$ in the weak coupling limit as suggested in the discussion leading to Eq. (15). For $\mathcal{X} > \mathcal{X}_c$, the theory is consistent with the classically predicted limit of $+6$, while for $\mathcal{X} < \mathcal{X}_c$, the limit $-3$ is seen. Similar results occur for non-zero values of $b$ as pictured in Figs. 2 and 4 where the results for $b = \pm 1$ are pictured. Again, $\mathcal{X}_c = -\frac{1}{3}$ emerges as a critical point separating two domains of convergence in the limit of weak coupling.

In (2+1) dimensions, the string tension as obtained from the area law behavior of Wilson loops can be shown to be a function of $\langle A_p \rangle_o$ alone [6] [17]. A sufficient condition for this result to hold is that the vacuum factorizes into pieces associated with each plaquette as it clearly does in this case. One finds that the expectation value for a loop enclosing an area $m$ measured in units of $a^2$ is given by the $m^{\text{th}}$ power of the fundamental plaquette. We then have,

$$a^2 K = -\frac{2}{3} \ln\left(\frac{\langle A_p \rangle_o}{6}\right). \tag{19}$$

The factor $\frac{2}{3}$ appears because the area law coming from spacelike Wilson loops corresponds to an octet (adjoint representation) color source and sink rather than the triplet (fundamental representation) $q$ - $\bar{q}$ potential that we really want [18].

In addition to the approach described above, we employ a more general method of extracting the area law behavior which is also applicable to the more complex (3+1)-d problem. Here we calculate the string tension directly from the expectation values of a sequence of increasingly larger square Wilson loops. When the loops are large (compared to a correlation length on the lattice), we expect the behavior to be dominated by area and perimeter laws,

$$\phi_I \approx \exp(-A - \frac{3}{2} K a^2 I^2 - C a 2I), \tag{20}$$

where $I$ indicates an $I$x$I$ loop and for fixed values of the lattice coupling, $A, K,$ and $C$ are constants. The factor $\frac{2}{3}$ appears as in Eq. (19). So, a quadratic least squares fit to the data,

$$-\ln(\phi_I) \approx A + \frac{3}{2}Ka^2I^2 + Ca2I, \qquad (21)$$

gives $A$, $\frac{3}{2}a^2K$, and $2aC$ for various values of $\xi$.

The asymptotic scaling behavior obtained from a perturbative expansion of the renormalization group function for the SU(3) theory in 2+1 dimensions is $a\xi =$ constant [19]. So, we will look for constant weak coupling behavior for lattice quantities such as string tension and glueball masses which are measured in units of the lattice spacing as an indication of consistency with the continuum theory.

Thus, we can see just from the results in Figs. 2,3,4 and Eq. (19) that only the theories corresponding to $\mathcal{X} > \mathcal{X}_c$ give meaningful string tension results.

We evaluate the integrals for the string tension calculation suggested by Eq. (21) using a Monte Carlo simulation on a finite 2-d space lattice. In the results section, we then compare string tension obtained in this way with the 'exact' results as a test of our implementation of the statistical integration technique and of the quadratic fitting approach to the string tension. This is of interest since in a subsequent paper, we will report results for the (3+1)-d problem where comparisons with exact results are not possible as they are here.

## IV. VARIATIONAL ESTIMATE OF SCALAR GLUEBALL SPECTRUM

Since the eigenvalues of $H$ in the continuum limit correspond to physical observables only when gauge invariant states are used, we begin by projecting a set of gauge invariant basis states orthogonal to the vacuum state,

$$|\psi_i\rangle = (\phi_i - \langle\phi_i\rangle_o)|\Psi_o\rangle, \quad \text{where} \quad \langle\phi_i\rangle_o \equiv \langle\Psi_o|\phi_i|\Psi_o\rangle. \qquad (22)$$

It is easy to demonstrate that the trace of any product of link variables constructed by traversing a closed path in the lattice is gauge invariant and is thus a potential basis function. As there are infinitely many candidates for $\phi_i$, an attempt must be made to choose a small subset of these which sufficiently spans the space occupied by the low-lying excitations of the theory. Obviously, the most efficient choice would be an orthogonal set of basis states.

As there appears to be no easy way to choose such a basis beforehand, we will work with a nonorthogonal basis and variationally minimize the excited state energies (M) by solving the generalized eigenvalue problem,

$$\det|\mathbf{H} - \frac{2aM}{g^2}\mathbf{O}| = 0, \tag{23}$$

with the definition of the Hermitian matrices

$$H_{ij} \equiv \frac{2a}{g^2}\langle\psi_i|H|\psi_j\rangle = -\langle[E_l^\alpha, \phi_i^\dagger][E_l^\alpha, \phi_j]\rangle_o$$
$$O_{ij} \equiv \langle\psi_i|\psi_j\rangle = \langle\phi_i^\dagger\phi_j\rangle_o - \langle\phi_i^\dagger\rangle_o\langle\phi_j\rangle_o. \tag{24}$$

Apart from the factor $\frac{2a}{g^2}$, the first matrix is just the Hamiltonian evaluated in the chosen basis and the second matrix gives the overlaps among the basis states and is a measure of the nonorthogonality of the basis. Thus, for a basis containing $n$ states, there are $n^2 + n$ integrals of the form,

$$\langle O(U_l)\rangle_o = \frac{\int O(U_l)\Psi_o^2(U_l)\{dU_l\}}{\int \Psi_o^2(U_l)\{dU_l\}}. \tag{25}$$

to evaluate. Evaluation of these matrix elements can be simplified in the same way as described for $\langle A_p\rangle_o$ in the previous section. However, we use the less efficient Monte Carlo approach since we are also interested in the (3+1)-d problem where the integrals can only be evaluated statistically.

The variational basis we choose consists of all square loops having up to 7 links to a side. They can be expressed as,

$$\phi_I = \frac{1}{L^2 N}\sum_{i=1}^{L^2}\mathrm{ReTr}(U_{Ii}), \tag{26}$$

where the lattice has $L$ links to a side and $U_{Ii}$ denotes the path-ordered product of link variables obtained by traversing the perimeter of an $I$x$I$ square loop located at lattice position $i$. Figure 5 gives a graphical representation of these 7 states. In the weak coupling (continuum) limit, where the rotational symmetries of continuous space are restored, this basis produces estimates of the $J^{\mathrm{PC}} = 0^{++}$ glueball spectrum.

## V. MONTE CARLO SIMULATION

Mont Carlo calculations were performed using a $21^2$ lattice having periodic boundary conditions [17] [20]. This lattice size is large enough to give finite size effects for the largest loops in our basis which are negligible compared to the statistical fluctuations in our Monte Carlo estimates. We determined this by repeating most of the computations presented in this paper with smaller $14^2$ and $7^2$ lattices. Differences between the $21^2$ and $14^2$ results were well within statistical noise. It was only with the $7^2$ lattice that large finite size effects were seen. Thus, we believe that the lattice used here is a very good approximation to an infinite one for the basis under consideration.

We looked at theories corresponding to the four parameter values, $\mathcal{X} = 0.5, 1.0, 1.5,$ and $2.0$ with $b = 0$ in each case. For each $\mathcal{X}$ we calculated the matrix elements of Eqs. (24) for 16 values of the coupling constant distributed over the range $0.30 \leq \xi \leq 2.00$. We employed a multi-hit Metropolis algorithm to generate an ensemble of quasi-independent lattice configurations, $\{U_{li}\}$ distributed according to the probability density,

$$\frac{\Psi_0^2(U_l)}{\int \Psi_0^2(U_l)\{dU_l\}} . \tag{27}$$

During the Metropolis sweeps of the lattice, each link was updated (hit) three times. The probability for accepting a link update was maintained near the value 0.5 by dynamically adjusting the step (SU(3) rotation) size. Thus, after each sweep, the probability for a link to remain unchanged is only $0.5^3 = 0.125$, increasing the independence of subsequent lattice configurations compared to a single hit approach. For each value of $\xi$, thermalization of a starting lattice configuration was followed by the generation of an ensemble of 84,000 lattice configurations partitioned into 20 subensembles of equal size. Typically, the starting lattice was taken from a prior simulation at a nearby value of the coupling in order to reduce the number of sweeps required to sufficiently approach the desired distribution.

Monte Carlo estimates of the expectation values are then simple ensemble averages,

$$\langle O(U_l)\rangle_0 = \frac{1}{N}\sum_{i=1}^{N} O(U_{li}) \pm \frac{\sigma_O}{\sqrt{N}} , \tag{28}$$

where $N = 84000$ in our case, and the estimate of the statistical error, $\sigma_O^2$, is calculated as the variance of estimates obtained from the 20 subensembles.

## VI. RESULTS

Vacuum expectation values of the square loops in our variational basis as functions of the lattice coupling $\xi$ are displayed in Fig. 6 for $\mathcal{X} = 0.5, 1.0, 1.5,$ and $2.0$ with $b = 0$. In the strong coupling limit, where $\xi \to 0$, the link variables can be expected to uncouple and distribute themselves uniformly over the SU(3) manifold. Our basis states, which are products of the links, must obey this same distribution and therefore have zero trace on average. The figure shows that the expectation values do approach zero in this limit. In weak coupling, where $\xi \to \infty$ and the correlation length diverges ($a \to 0$), neighboring links should become increasingly correlated. In this limit, the normalized traces of Eq. (26) approach unity for all $I$. The figure shows how, as the correlation length increases, so do the loop expectation values. Our basis states are becoming more and more alike and the off-diagonal elements of the overlap matrix continue to grow in size as the lattice spacing shrinks. If $\xi$ were taken large enough, numerical singularity of the overlap matrix would eventually occur.

At a given value of the lattice coupling, Fig. 6 demonstrates that the links become more highly correlated as $\mathcal{X}$ increases. So the approach to continuum behavior of the system (with increasing $\xi$) appears to be more rapid for larger values of $\mathcal{X}$.

The area law behavior of these 7 states was extracted from the Monte Carlo data both by least squares fitting as suggested in Eq. (21) and from the plaquette expectation value as in Eq. (19). The resulting string tensions are displayed in Fig. 7 along with the exact result. Agreement between Monte Carlo and the exact result is nearly perfect except for some significant statistical errors in the least squares result in the strong coupling region. Here, the larger loops have large fractional uncertainties (because their numerical values are close to zero) and the least squares method requires at least three well determined data points. Some least squares estimates at $\xi = 0.30$ and $0.40$ are missing because the data was not accurate enough for the fitting routines to function properly.

For all four values of $\mathcal{X}$, the resulting string tensions agree both in the strong and weak coupling limits. It is only in the transition between the two regimes that the theories differ. They appear to be approaching a common constant value of $a\sqrt{K}\xi = 0.256$ in the continuum limit. Of the four $\mathcal{X}$ values considered, $\mathcal{X} = 1.5$ shows the strongest appraoch to continuum behavior becoming constant near $\xi = 0.4$.

Next, we diagonalize our Hamiltonians within the space defined by the variational basis. Masses associated with the four lowest eigenstates along with the mass-to-string tension ratio are displayed in Figs. 8 through 11 for each of the four values of $\mathcal{X}$ we considered. Errorbars are included in these figures but are nearly invisible since they are small compared to the size of the symbols and the vertical scale of the plots.

The mass gap (lowest eigenstate) enjoys a rather large scaling window extending approximately over $0.40 \leq \xi \leq 1.4$ for $\mathcal{X} = 2.0$. This is to be contrasted with the result for $\mathcal{X} = 0.5$ where scaling set in only for $\xi \geq 0.8$. Again we see that larger $\mathcal{X}$ results in continuum behavior at smaller values of $\xi$. For $\xi \geq 1.4$, basis truncation causes deviation from scaling since the decreasing lattice spacing implies that progressively larger loops must be included in the variational basis in order to accurately represent the glueballs. The mass-to-string tension ratio curves are similar to the mass curves except that the differences in the scaling region among the four $\mathcal{X}$ values is enhanced by the slower convergence of the string tension for small $\mathcal{X}$.

Results for the second lowest eigenstate are very similar to those obtained for the mass gap except that the scaling windows extend only to about $\xi = 1.0$. This is to be expected since excited states should be physically larger than the ground state (as in other familiar bound state problems such as the hydrogen atom) and should then suffer from the limitations of the basis truncation at smaller values of $\xi$ than in the ground state case. In fact, for $\mathcal{X} = 0.5$, the second eigenstate fails to scale at all. An approach to scaling can be seen for the third eigenstate in the region $0.6 \leq \xi \leq 0.8$ for the three $\mathcal{X}$ values larger than $0.5$.

Finally, the fourth eigenstate seems to be beyond what our limited basis can accurately represent.

Examining Figs. 10 and 11 reveals an interesting trade off with regard to choosing the 'best' value for the parameter $\mathcal{X}$. It is due to the fact that the variational basis is different for different values of $\mathcal{X}$. Recall that the set of projectors, $\phi_i$, are the same in every case but that the basis consists of the parts which are orthogonal to the vacuum which is a function of $\mathcal{X}$. So, the question then becomes which $\mathcal{X}$ will give the best basis for representing the glueballs. Our results show that although larger $\mathcal{X}$ produces better scaling in the sense that it occurs over a larger window in $\xi$, one appears to pay for this by getting somewhat poorer variational estimates of the excited state energies. For example, the fourth eigenstate shows approximately a 10% difference between the energies coming from the theories corresponding to $\mathcal{X} = 0.5$ and $\mathcal{X} = 2.0$. The lowest eigenstate seems to be unaffected by the modification of the basis for different $\mathcal{X}$; that is, the theories agree on the size of the mass gap.

## VII. SUMMARY

The results obtained here show that there is a class of lattice Hamiltonians for SU(3) gauge theory which, in spite of having a range of different classical limits, have the same scaling limit for the $J^{PC} = 0^{++}$ glueball masses and string tension. This universality holds provided that the vacuum expectation value of a plaquette also has the correct limit, which in the present calculations occurs for $\mathcal{X} > -\frac{1}{3}$. Calculations using plaquette space integration [21] show that similar conclusions hold for SU(2) as well as for SU(3) so we anticipate it is true for all N.

Due to an error, previous results in this field [3] did not show true universality in (2+1)-dimensions, nor was it realized that there is a continuous class of Hamiltonians (i.e. any value of $\mathcal{X} > -\frac{1}{3}$) which yields universality. The present work allows us to set the stage for (3+1) dimensional lattice calculations with the knowledge that this class of lattice Hamiltonians in (3+1) will have the same classical continuum limit as the Kogut-Susskind Hamiltonian. The latter is already known to have the desired classical continuum limit as QCD. Calculations are in progress using Monte Carlo techniques in (3+1) dimensions for SU(3) and will be reported on later.

### APPENDIX: CLASSICAL CONTINUUM LIMIT

In this appendix we examine the (2+1)-d classical continuum limit of the vacuum fixing term, $\Delta H$ given in Eq. (10),

$$\Delta H = -\frac{e^2}{2}(E_l^\alpha E_l^\alpha R) - \frac{e^2}{2}(E_l^\alpha R)(E_l^\alpha R) + \frac{1}{a^2 e^2}\sum_p A_p \,, \tag{A1}$$

where the form of $R$ is defined by Eq. (7). Using Eqs. (7), (8), and (2) one can easily verify that,

$$E_l^\alpha R = \sum_{p \supset l}(\beta_1 + 2\beta_2 A_p)(E_l^\alpha A_p) \,, \tag{A2}$$

and

$$E_l^\alpha E_l^\alpha R = \sum_{p \supset l}[(\beta_1 + 2\beta_2 A_p)(E_l^\alpha E_l^\alpha A_p) + 2\beta_2(E_l^\alpha A_p)(E_l^\alpha A_p)] \,, \tag{A3}$$

where $\sum_{p \supset l}$ is a sum restricted to plaquettes sharing the link $l$. $\Delta H$ can be separated into the following three terms,

$$\begin{aligned}
\Delta H_1 &= \frac{e^2}{2}\sum_l \sum_{p \supset l} C_N(\beta_1 + 2\beta_2 A_p)A_p - \frac{1}{a^2 e^2}\sum_p A_p \\
\Delta H_2 &= \frac{e^2}{2}\sum_l \sum_{p \supset l} 2\beta_2(E_l^\alpha A_p)(E_l^\alpha A_p) \\
\Delta H_3 &= \frac{e^2}{2}\sum_l [\sum_{p \supset l}(\beta_1 + 2\beta_2 A_p)(E_l^\alpha A_p)]^2 \,,
\end{aligned} \tag{A4}$$

where $C_N = \frac{1}{4}\lambda^\alpha \lambda^\alpha = \frac{N^2-1}{2N}$ in the expression for $\Delta H_1$ is the Casimir invariant. We will now take each of these terms in turn and examine the behavior as $a \to 0$.

**1. $\Delta H_1$**

Since $\text{Tr} U_p$ is invariant under cyclic permutations of the four link variables from which $U_p$ is constructed, the double sum over any function of this trace, $\sum_l \sum_{p \supset l} f(\text{Tr} U_p)$, is just four times the sum over plaquettes, $4 \sum_p f(\text{Tr} U_p)$. $\Delta H_1$ of Eqs. (A4) can then be expressed as a simple plaquette sum,

$$\Delta H_1 = 2e^2 C_N \sum_p (\beta_1 + 2\beta_2 A_p - \frac{1}{2 C_N a^2 e^4}) A_p . \tag{A5}$$

The function $A_p$ has a well known continuum limit which is obtained by making a Taylor series expansion of the link functions and collecting like powers of the lattice spacing. The result is,

$$A_p \xrightarrow{a \to 0} 2N - \frac{1}{2} e^2 a^4 G^\alpha_{12}(\vec{x}) G^\alpha_{12}(\vec{x}) + O(a^8) . \tag{A6}$$

The eight gauge field tensors associated with a plaquette in the 12 plane at lattice location $\vec{x}$ are defined as,

$$G^\alpha_{12}(\vec{x}) = \partial_1 A^\alpha_2(\vec{x}) - \partial_2 A^\alpha_1(\vec{x}) - e f^{\alpha\beta\gamma} A^\beta_1(\vec{x}) A^\gamma_2(\vec{x}) . \tag{A7}$$

Here, $A^\alpha_i(\vec{x})$ denote the two spatial components of the $N^2 - 1$ gauge potentials, and $f^{\alpha\beta\gamma}$ are structure constants defined via commutation relations among the SU(N) group generators,

$$[\frac{\lambda^\alpha}{2}, \frac{\lambda^\beta}{2}] = i f^{\alpha\beta\gamma} \frac{\lambda^\gamma}{2} . \tag{A8}$$

Using Eq. (A6) together with $\sum_p a^2 \to \int d^2 x$, the leading terms in $\Delta H_1$ are,

$$\Delta H_1 = \frac{2e^2 C_N}{a^2} \int d^2 x (2N - \frac{1}{2} e^2 a^4 G^\alpha_{12}(\vec{x}) G^\alpha_{12}(\vec{x}) + O(a^8))$$
$$(\beta_1 + 4N\beta_2 - \frac{1}{2 C_N a^2 e^4} - \beta_2 e^2 a^4 G^\alpha_{12}(\vec{x}) G^\alpha_{12}(\vec{x}) + O(a^6)) . \tag{A9}$$

All terms greater than $O(a^0)$ can be eliminated by choosing the parameters $\beta_1$ and $\beta_2$ such that,

$$\beta_1 + 4N\beta_2 = \frac{1}{2 C_N a^2 e^4} , \tag{A10}$$

which gives,

$$\Delta H_1 = \beta_2[4NC_N a^2 e^4 \int d^2x G^\alpha_{12}(\vec{x})G^\alpha_{12}(\vec{x}) + O(a^6)] . \tag{A11}$$

Notice first the above constraint does not determine the $a$ dependence of $\beta_1$ and $\beta_2$. The only 'additional' restriction on the form of the $\beta$'s is that the left hand side of Eq. (A10) should remain always positive so that the lattice coupling is real. We see then that the classical limit of our Hamiltonian is not fixed by Eq. (A10) alone; one must further specify how $\beta_1$ and $\beta_2$ individually depend on $a$. The possibilities fall into three catagories. First if $\beta_2 < O(a^{-2})$, then $\Delta H_1$ vanishes. This can be realized by choosing $\beta_2 =$constant, for example. Secondly, in the event that $\beta_2 > O(a^{-2})$, $\Delta H_1$ is divergent for small $a$. Finally, when $\beta_2$ is exactly of order $a^{-2}$, $\Delta H_1$ has a finite non-zero limit which depends on the precise form chosen. As an example we may put $\beta_1 = \frac{\mathcal{X}}{2C_N a^2 e^4}$ and $\beta_2 = \frac{1-\mathcal{X}}{8NC_N a^2 e^4}$ in which case the classical limit for $\Delta H_1$ becomes,

$$\begin{aligned}\Delta H_1 &= \frac{1-\mathcal{X}}{2} \int d^2x G^\alpha_{12}(\vec{x})G^\alpha_{12}(\vec{x}) + O(a^4) \\ &= (1-\mathcal{X}) \int d^2x \frac{1}{2}\vec{B}^\alpha \cdot \vec{B}^\alpha .\end{aligned} \tag{A12}$$

We write the color-magnetic field as a vector to make clear the connection with the more familiar case of 3 space dimensions. In (2+1)-d, $B^\alpha$ has only one component. We see that $\mathcal{X} \neq 1$ implies a modification of the magnetic part of the Hamiltonian relative to the standard Kogut-Susskind form. Later in this section, we discuss the implications of this result.

### 2. $\Delta H_2$

Since $\text{Tr}(E^\alpha_l A_p) \neq \text{Tr}(E^\alpha_{l'} A_p)$ in general for $l \neq l' \subset p$, we can not directly make the replacement $\sum_l \sum_{p \supset l} \to 4\sum_p$ as we did in the analysis of $\Delta H_1$. Instead, we let $U_1, U_2, U_3$, and $U_4$ be the four links which define a particular plaquette and define,

$$U_{p1} = U_1 U_2 U_3^\dagger U_4^\dagger$$

$$U_{p2} = U_2 U_3^\dagger U_4^\dagger U_1$$

$$U_{p3} = U_3 U_2^\dagger U_1^\dagger U_4$$

$$U_{p4} = U_4 U_3 U_2^\dagger U_1^\dagger \ . \tag{A13}$$

Now the double sum in $\Delta H_2$ from Eqs. (A4) can be expressed as an unrestricted sum over plaquettes,

$$\Delta H_2 = e^2 \sum_p \beta_2 \sum_{i=1}^{4} \{E_i^\alpha \text{Tr}[U_{pi} + U_{pi}^\dagger]\}^2$$

$$= e^2 \sum_p \beta_2 \sum_{i=1}^{4} \{\text{Tr}[\frac{\lambda^\alpha}{2}(U_{pi} - U_{pi}^\dagger)]\}^2 \ , \tag{A14}$$

where we have also used the definition of $A_p$ and Eqs. (2). When $a$ is small the trace appearing in Eq. (A14) can be expressed,

$$\text{Tr}[\frac{\lambda^\alpha}{2}(U_{pi} - U_{pi}^\dagger)] = \text{Tr}[\frac{\lambda^\alpha}{2}(\pm 2iea^2 G_{12}^{\alpha'} \frac{\lambda^{\alpha'}}{2} + O(a^6))]$$

$$= \pm 2iea^2 G_{12}^{\alpha'} \text{Tr}[\frac{\lambda^\alpha}{2}\frac{\lambda^{\alpha'}}{2}] + O(a^6)$$

$$= \pm iea^2 G_{12}^\alpha + O(a^6) \ , \tag{A15}$$

where the sign $\pm$ depends on the sense of rotation in the definition of $U_{pi}$. In Eqs. (A13), $U_{p1}$ and $U_{p2}$ involve $G_{12}$ whereas $U_{p3}$ and $U_{p4}$ circulate in the opposite sense and involve $G_{21} = -G_{12}$. In the final step above we used the normalization of the generators, $\text{Tr}(\frac{\lambda^\alpha}{2}\frac{\lambda^{\alpha'}}{2}) = \frac{\delta_{\alpha\alpha'}}{2}$. The leading term of Eq. (A14) can now be written,

$$\Delta H_2 = e^2 \beta_2 \sum_p \sum_{i=1}^{4} \{\pm iea^2 G_{12}^\alpha\}^2$$

$$= -4e^4 a^2 \beta_2 \int d^2 x G_{12}^\alpha(\vec{x}) G_{12}^\alpha(\vec{x}) \ , \tag{A16}$$

which comparing to Eq. (A11) is proportional to the limit for $\Delta H_1$,

$$\Delta H_2 = -\frac{1}{NC_N}\Delta H_1 \ , \text{ as } a \to 0$$

$$= -\frac{1}{4}\Delta H_1 \ , \text{ SU(3)} \ . \tag{A17}$$

### 3. $\Delta H_3$

Finally, consider $\Delta H_3$. In order to decide its limit, let $p'$ and $p''$ be the two plaquettes which share a common link $l$. Furthermore, we define $U_{p'}$ and $U_{p''}$ such that $l$ is the first link occuring in the product of four. That is, $U_{p'} = U_l U_m U_n^\dagger U_k^\dagger$, and similarly for $U_{p''}$. The expression for $\Delta H_3$ appearing in Eqs. (A4) then becomes,

$$\Delta H_3 = \frac{e^2}{2}\sum_l\{(\beta_1 + 2\beta_2 A_{p'})(E_l^\alpha A_{p'}) + (\beta_1 + 2\beta_2 A_{p''})(E_l^\alpha A_{p''})\}^2$$
$$= \frac{e^2}{2}\sum_l\{(\beta_1 + 2\beta_2 A_{p'})\text{Tr}[\frac{\lambda^\alpha}{2}(U_{p'} - U_{p'}^\dagger)] + (\beta_1 + 2\beta_2 A_{p''})\text{Tr}[\frac{\lambda^\alpha}{2}(U_{p''} - U_{p''}^\dagger)]\}^2 . \quad \text{(A18)}$$

Using the result of Eq. (A15) we have,

$$\text{Tr}[\frac{\lambda^\alpha}{2}(U_{p'} - U_{p'}^\dagger)] = iea^2 G_{12}^\alpha(\vec{x}') + O(a^6) . \quad \text{(A19)}$$

and

$$\text{Tr}[\frac{\lambda^\alpha}{2}(U_{p''} - U_{p''}^\dagger)] = iea^2 G_{21}^\alpha(\vec{x}'') + O(a^6) . \quad \text{(A20)}$$

The relative ordering of the tensor indices in the two expressions above (12 vs. 21) occurs because $p'$ and $p''$ are defined such that they have opposite rotational sense. $\Delta H_3$ can now be written as,

$$\Delta H_3 = -\frac{e^4 a^4}{2}\sum_l\{(\beta_1 + 4N\beta_2)[G_{12}^\alpha(\vec{x}') + G_{ji}^\alpha(\vec{x}'')] + O(a^6)\}^2$$
$$= -\frac{1}{8C_N^2 e^4}\sum_l\{[G_{12}^\alpha(\vec{x}') + G_{ji}^\alpha(\vec{x}'')] + O(a^6)\}^2$$
(A21)

where in the second step we have used Eq. (A10). In the continuum limit, $\vec{x}'' = \vec{x}' + d\vec{x}$ and the two field tensors appearing in the brackets must have a relative minus sign between them from antisymmetry. It follows then that this sum is just the spatial derivative with the appropriate power of $a$. Futhermore, the sum over lattice links can be replaced by an area integral together with a sum over the 2 spatial coordinates. That is, we can use $\sum_l a^2 \to \int d^2 x \sum_{i=x,y}$. The result is,

$$\Delta H_3 = -\frac{1}{8C_N^2 e^4} \int d^2x [(\partial_i G_{12}^\alpha(\vec{x}))^2 + (\partial_j G_{12}^\alpha(\vec{x}))^2]$$
$$= -\frac{1}{8C_N^2 e^4} \int d^2x [(\nabla \times \vec{B}^\alpha) \cdot (\nabla \times \vec{B}^\alpha)] . \quad (A22)$$

As in the case of $\Delta H_1$, we have written the vector form for $B^\alpha$ even though in the current formulation it has only one component. The most significant feature of this result is that for any choice of the $\beta$'s subject to the constraint given in Eq. (A10), $\Delta H_3$ makes the *same* contribution to the Hamiltonian.

### 4. result

In (2+1) dimensions, the modified Hamiltonian under consideration has a classical continuum limit which differs from the standard Kogut-Susskind limit in the following way. Recall that the modified Hamiltonian is given by $H = H_{\text{KS}} - \Delta H_1 - \Delta H_2 - \Delta H_3$. Based on the preceeding analysis of these three terms, the classical limit for this Hamiltonian is dependent on the exact form chosen for the $\beta$'s. If we decide that we want a finite limit, then Eq. (A10) along with the condition $\beta_2 \leq O(a^{-2})$ must be satisfied. An interesting choice which obeys these conditions was given in terms of the parameter $\mathcal{X}$. With this choice we find that the magnetic term is modified relative to the Kogut-Susskind form in a way which depends on $\mathcal{X}$,

$$H \xrightarrow{a \to 0} \frac{1}{2} \int (\vec{E}^\alpha \cdot \vec{E}^\alpha + \frac{1+3\mathcal{X}}{4} \vec{B}^\alpha \cdot \vec{B}^\alpha) d^2x + Q , \quad (A23)$$

where Q is shorthand for the $\mathcal{X}$-independent result for $\Delta H_3$ given by Eq. (A22). We see that the critical value $\mathcal{X}_c = -\frac{1}{3}$ marks the boundary between two regions where the magnetic contributions come in with opposite signs.

### APPENDIX: ACKNOWLEDGMENTS

The authors wish to thank Chris Long for his many helpful suggestions during the preparation of the manuscript. This research was supported by the Florida State University

Department of Physics, the Supercomputer Computations Research Institute, the Kansas Institute for Theoretical and Computational Science through the K*STAR/NSF program, and by the U.S. Department of Energy contract DE-FG02-86ER40273.24

# REFERENCES

[1] K. Wilson, Phys. Rev. **D10** (1974), 2445.

[2] J. Kogut and L. Susskind, Phys. Rev. **D11** (1975), 395.

[3] G. Shuohong, Z. Weihong, and L. Jinming, Phys. Rev. **D38** (1988), 2591.

[4] G. Shuohong and Z. Weihong, Phys. Rev. **D39** (1989), 3144.

[5] S. A. Chin, C. Long, and D. Robson, Phys. Rev. **D37** (1988), 3001.

[6] H. Arisue, M. Kato, and T. Fujiwara, Prog. Th. Phys. **70** (1983), 229.

[7] H. Arisue, Prog. Th. Phys. **84** (1990), 951.

[8] S. A. Chin, C. Long, and D. Robson, Phys. Rev. Lett. **57** (1986), 2779.

[9] C. Long, D. Robson, and S. A. Chin, Phys. Rev. **D37** (1988), 3006.

[10] R. Z. Roskies, Phys. Rev. **D39** (1989), 3177.

[11] J. Greensite, Phys. Lett. **191B** (1987), 431.

[12] J. Greensite and J. Iwasaki, Phys. Lett. **223B** (1989), 207.

[13] H. Arisue, Phys. Lett. **B280** (1992), 85.

[14] M. Teper, Phys. Lett. **B289** (1992), 115.

[15] M. Teper, Phys. Lett. **B311** (1993), 223.

[16] *Variational Study of SU(2) Hamiltonian Lattice Gauge Theory Using Both Analytical and Biased-Selection Monte Carlo Integration Techniques*, M. C. Huang, dissertation University of Florida (1988).

[17] *Variational Estimate Of Pure Gauge Field QCD Using a Lattice Hamiltonian With Exact Vacuum State*, G. M. Frichter, dissertation Florida State University (1993).

[18] N. A. Campbell, I. H. Joryz and C. Michael, Phys. Lett. **167B**, 91 (1986).

[19] B. A. Berg and A. H. Billoire, Nucl. Phys. **B230** (1984), 49.

[20] N. Metropolis, A. W. Rosenbluth, M. N. Rosenbluth, A. H. Teller, and E. Teller, Phys. Rev. **21** (1953), 1087.

[21] D. Robson, C. Long and S. A. Chin, in *Relativistic Many Body Physics*, B. C. Clark, R. J. Perry and J. P. Vary editors, World Scientific (1989).

# FIGURES

FIG. 1. Relationships among variables used to specify the Hamiltonian. To simplify the picture we consider here the case $b = 0$. The two darker grey areas indicate regions with distinct weak coupling ($\xi \to \infty$) limits for the quantum system. For example, $\mathcal{X} > \frac{1}{3}$ and $\mathcal{X} < \frac{1}{3}$ gives $+6$ and $-3$ resectively for the vacuum expectation of $A_p$ as the lattice cutoff is removed.

FIG. 2. Vacuum expectation value of the plaquette, $\langle A_p \rangle_o$, as a function of $\mathcal{X}$ and the lattice coupling, $\xi$ with $b = -1$. In weak coupling, $\mathcal{X}_c = -\frac{1}{3}$ marks the boundary between two domains of convergence.

FIG. 3. Vacuum expectation value of the plaquette, $\langle A_p \rangle_o$, as a function of $\mathcal{X}$ and the lattice coupling, $\xi$ with $b = 0$. In weak coupling, $\mathcal{X}_c = -\frac{1}{3}$ marks the boundary between two domains of convergence.

FIG. 4. Vacuum expectation value of the plaquette, $\langle A_p \rangle_o$, as a function of $\mathcal{X}$ and the lattice coupling, $\xi$ with $b = 1$. In weak coupling, $\mathcal{X}_c = -\frac{1}{3}$ marks the boundary between two domains of convergence.

FIG. 5. Projectors for the variational basis used in the $0^{++}$ glueball calculation.

FIG. 6. Monte Carlo results for the vacuum expectation values of the seven Wilson loop operators defining the variational basis. Different curve types correspond to four different values of $\mathcal{X}$.

FIG. 7. SU(3) string tension calculated from Eq.(15) as a function of $\mathcal{X}$ (see text) and the lattice coupling, $\xi$. All theories corresponding to $\mathcal{X} > \mathcal{X}_c$ yield the same string tension in the continuum limit although the rate of convergence differs.

FIG. 8. Lowest eigenstate and its ratio with respect to string tension as a function of the lattice coupling for $\mathcal{X} = 0.5, 1.0, 1.5,$ and $2.0$.

FIG. 9. Second lowest eigenstate and its ratio with respect to string tension as a function of the lattice coupling for $\mathcal{X} = 0.5, 1.0, 1.5,$ and $2.0$.

FIG. 10. Third lowest eigenstate and its ratio with respect to string tension as a function of the lattice coupling for $\mathcal{X} = 0.5, 1.0, 1.5,$ and $2.0$.

FIG. 11. Fourth lowest eigenstate and its ratio with respect to string tension as a function of the lattice coupling for $\mathcal{X} = 0.5, 1.0, 1.5,$ and $2.0$.

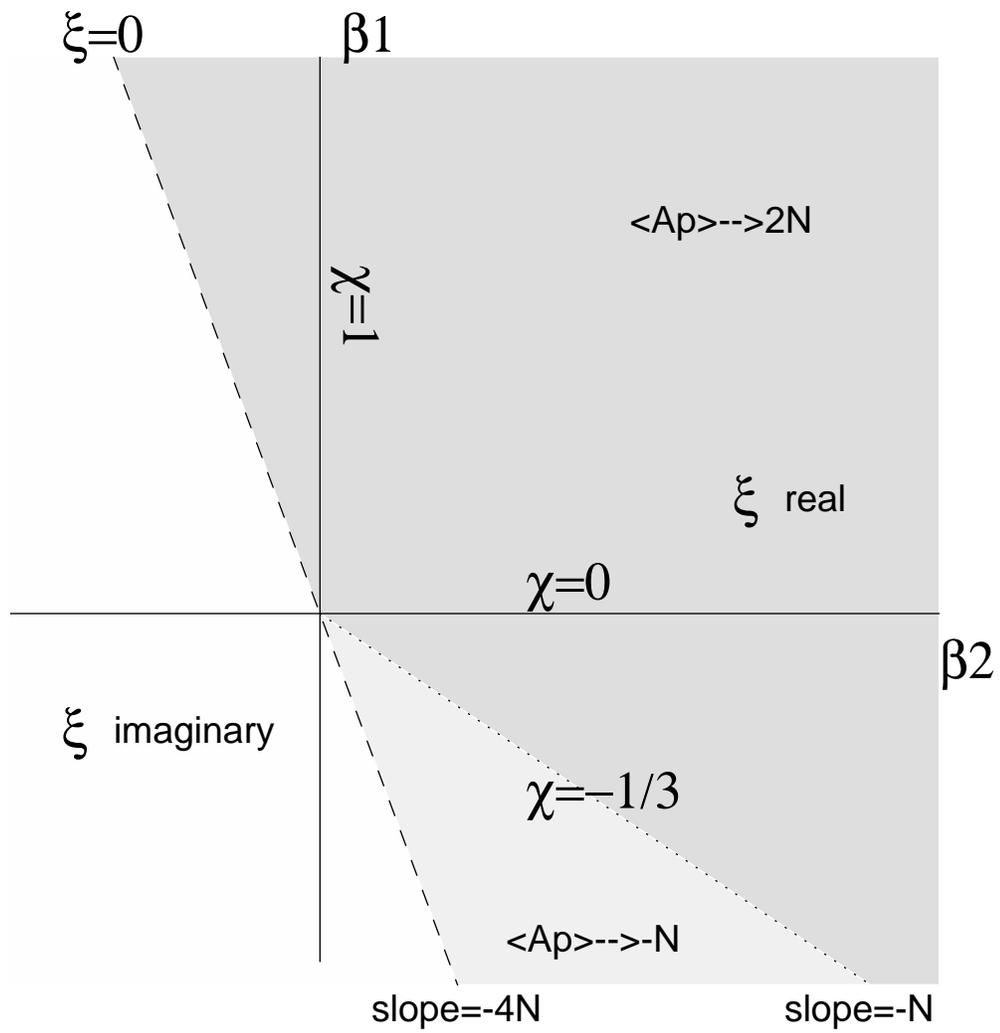

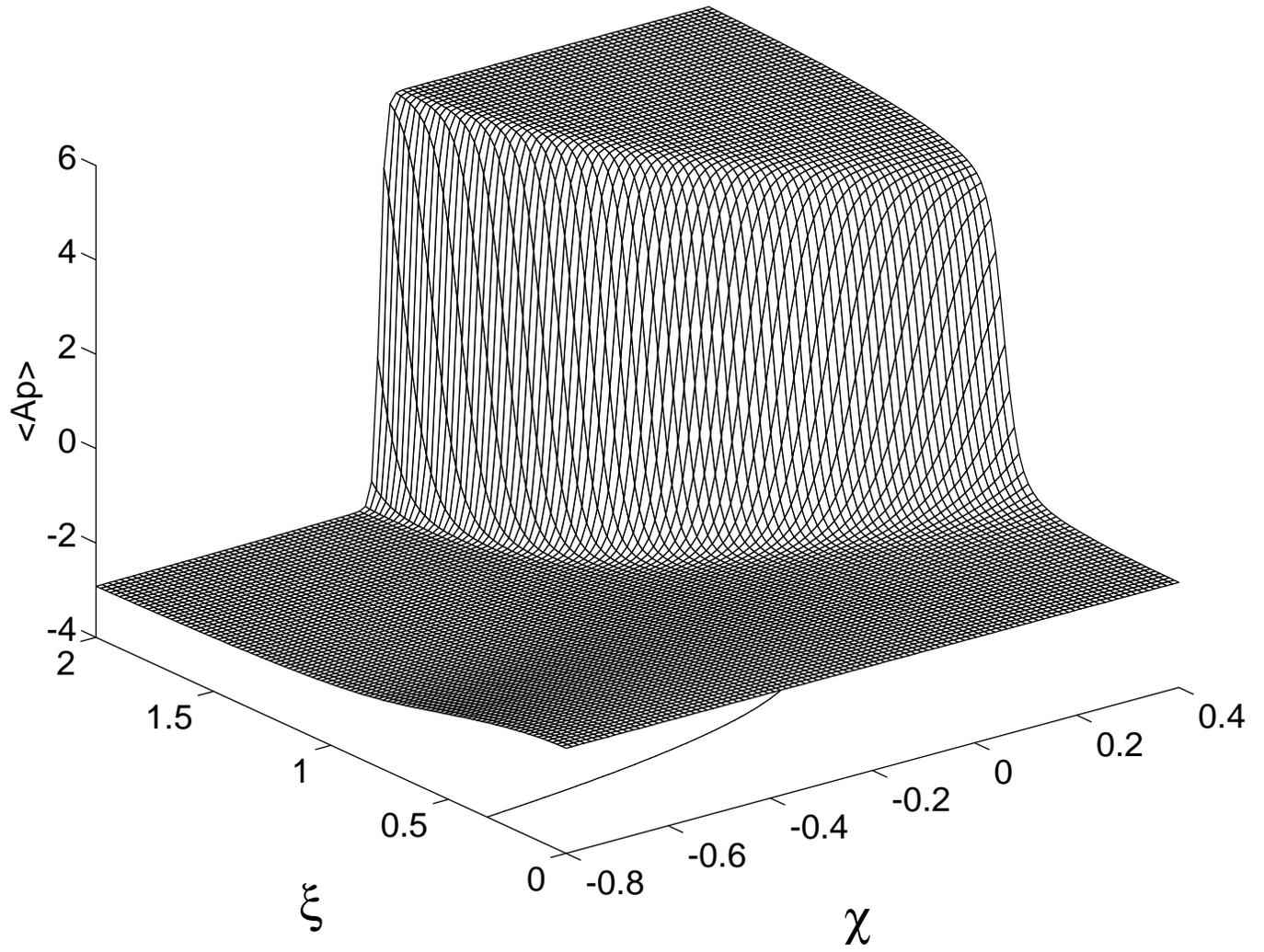

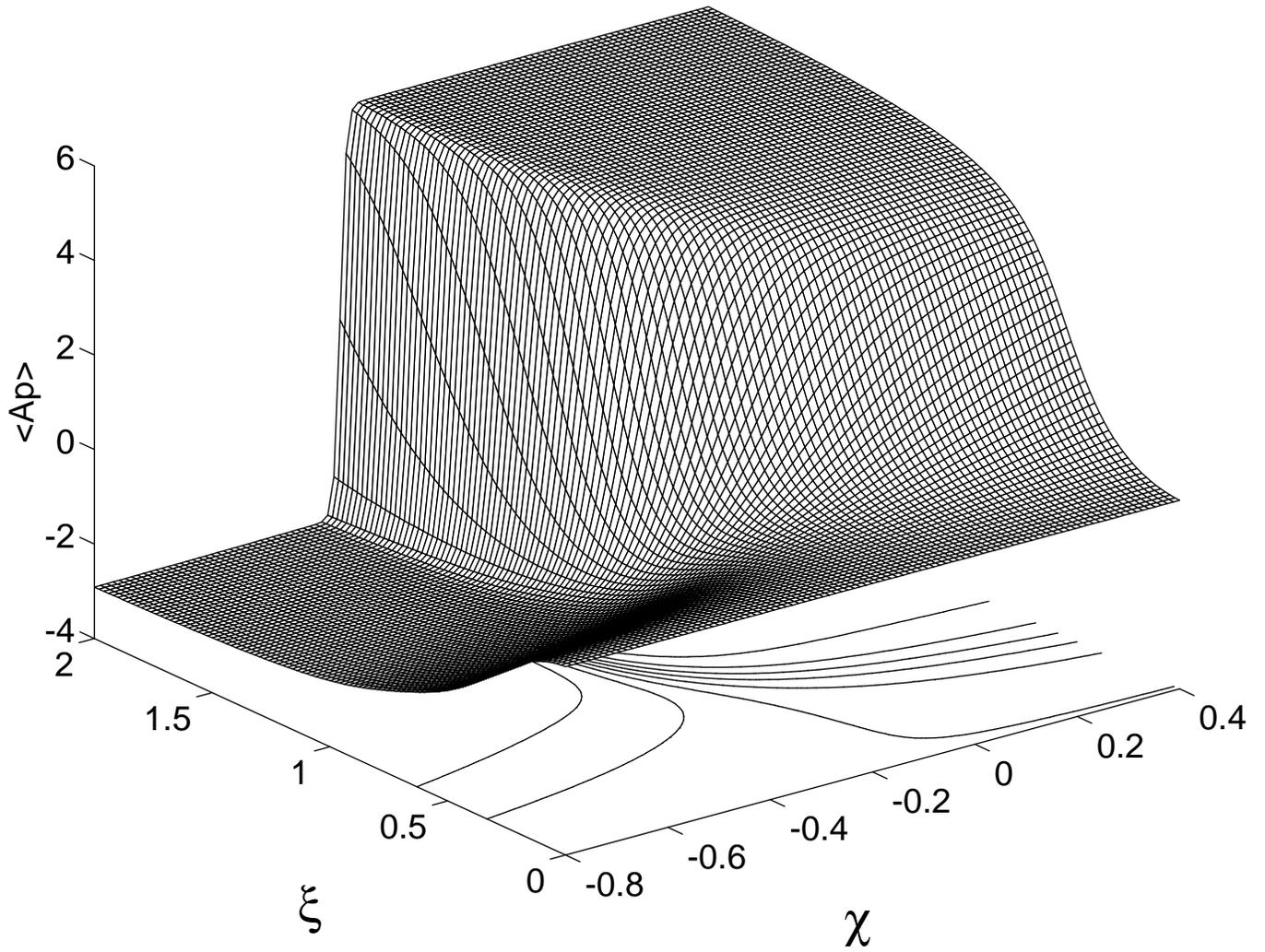

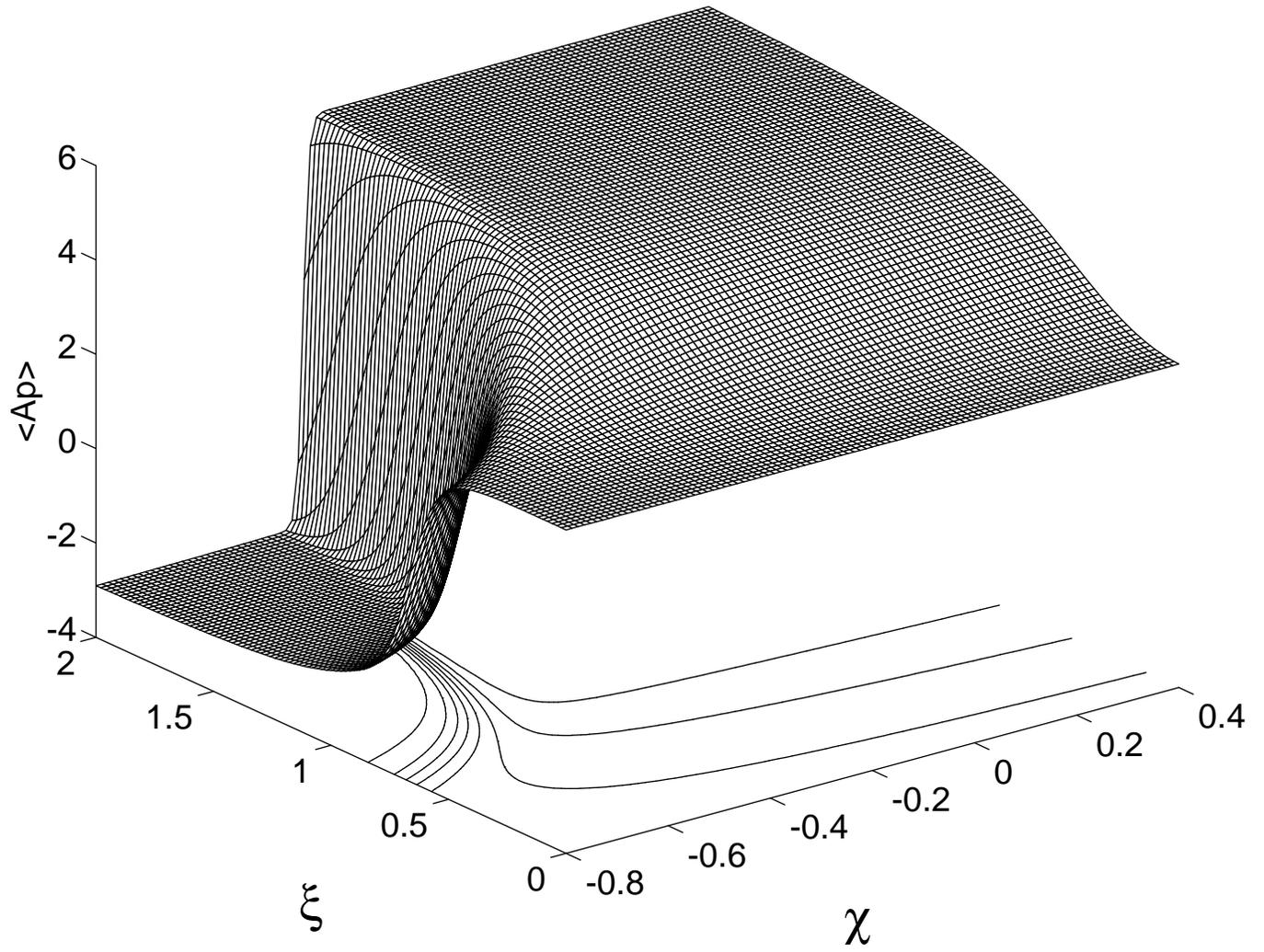

1 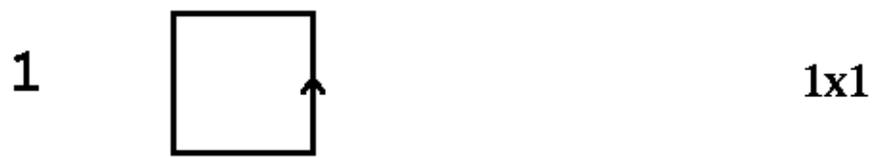 1x1

2 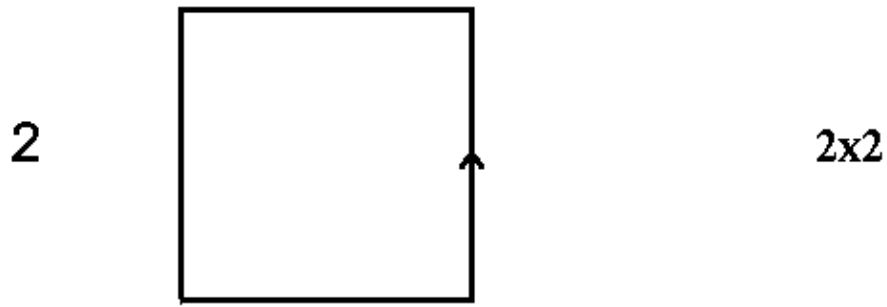 2x2

3 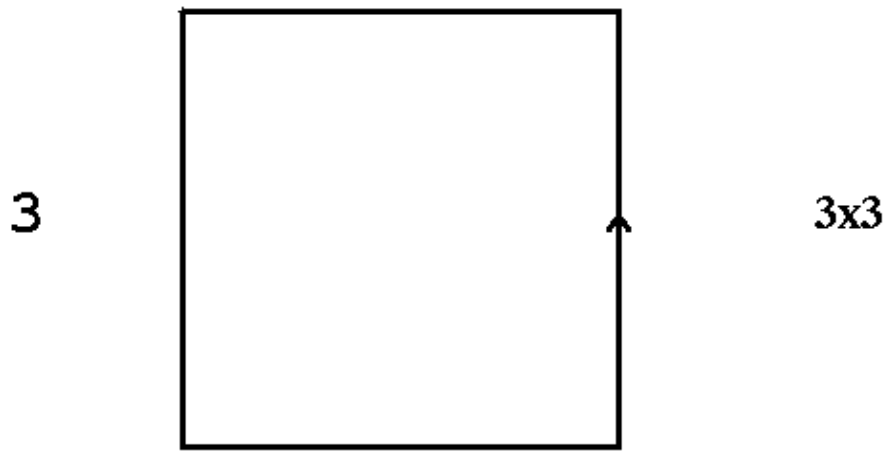 3x3

.
.
.
.

and so on thru 7x7

⟨ | x | ⟩ vs ξ

solid   χ=0.5
dot     χ=1.0
dash    χ=1.5
dash-dot χ=2.0

l=1, l=2, l=3, l=4, l=5, l=6, l=7

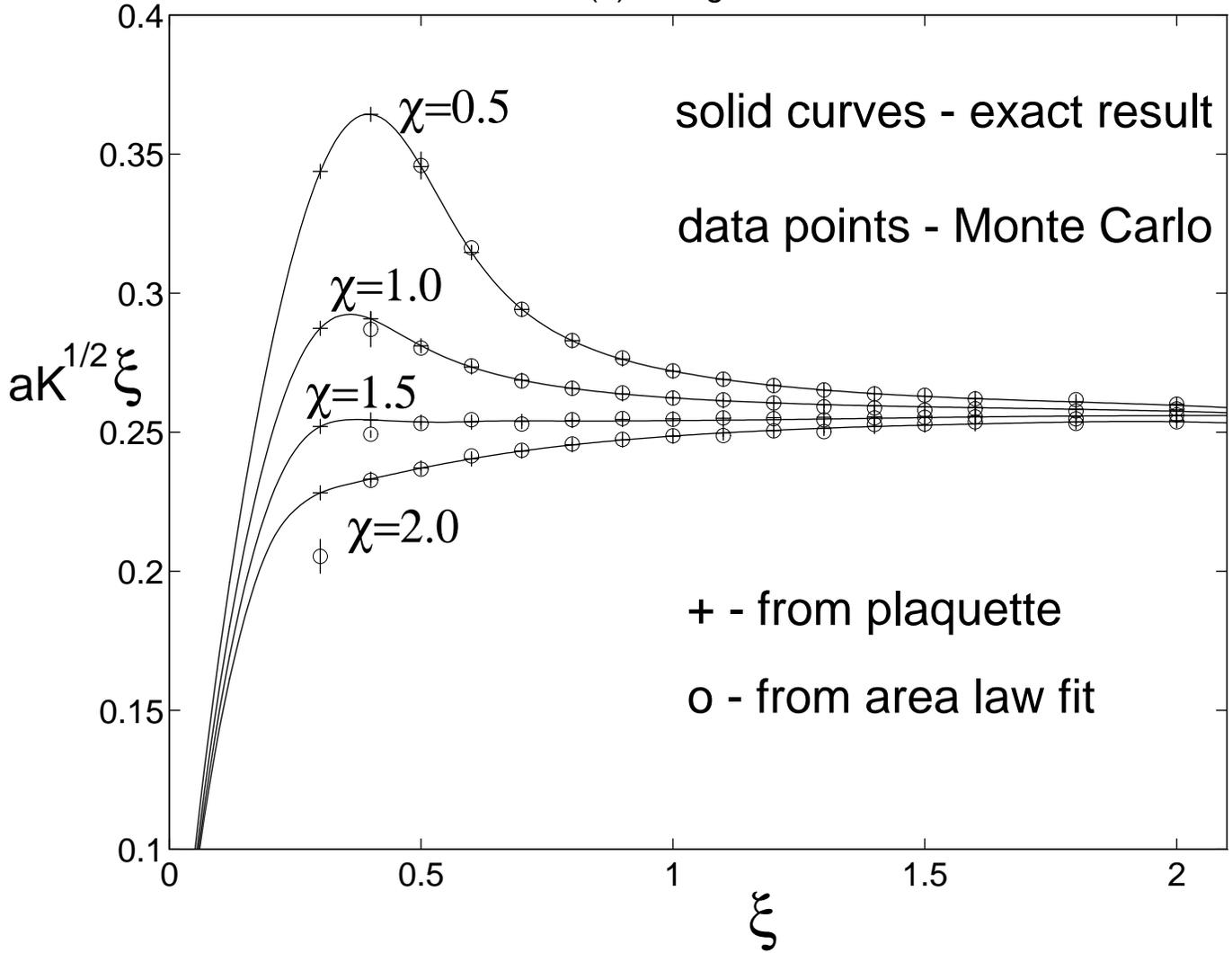

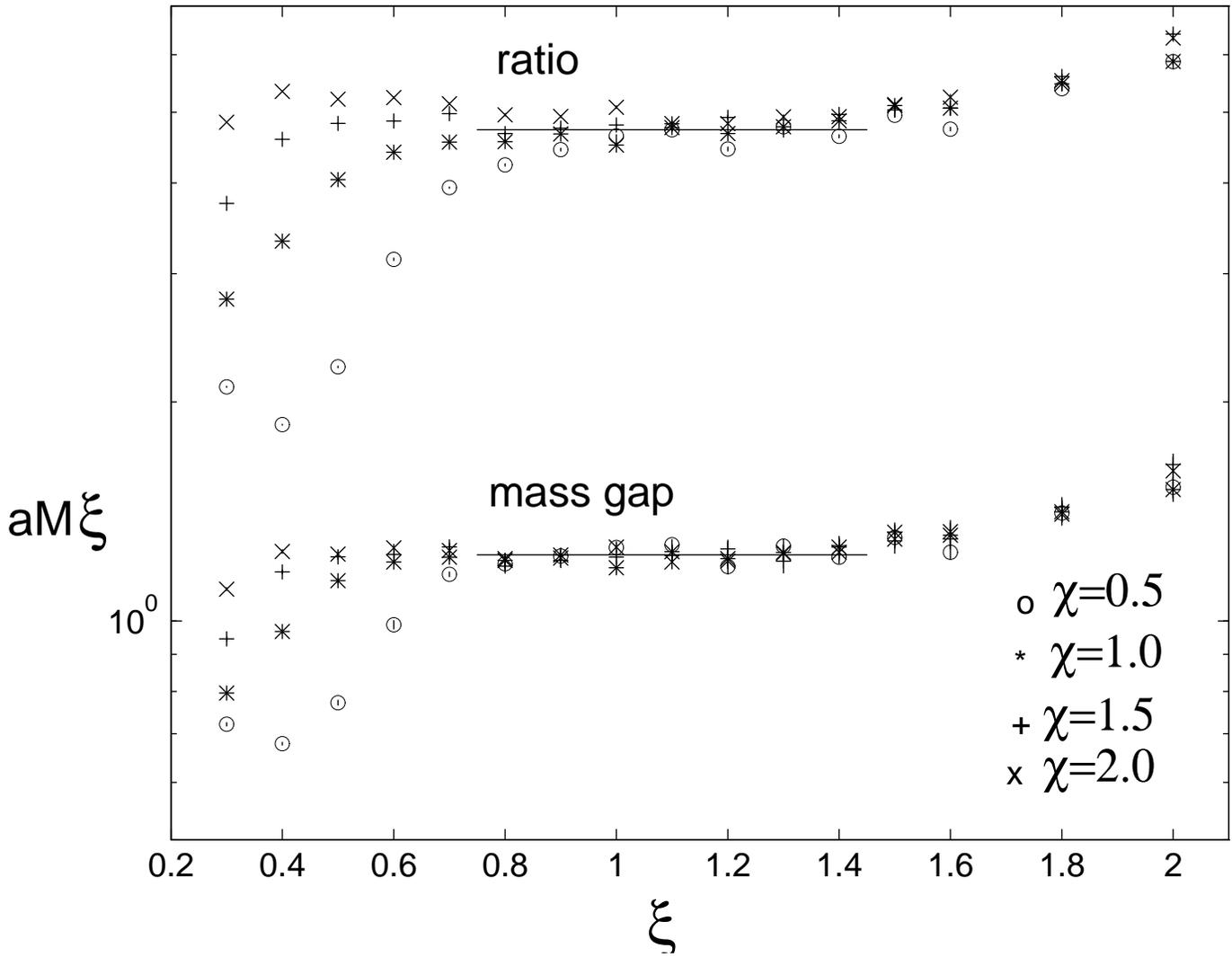

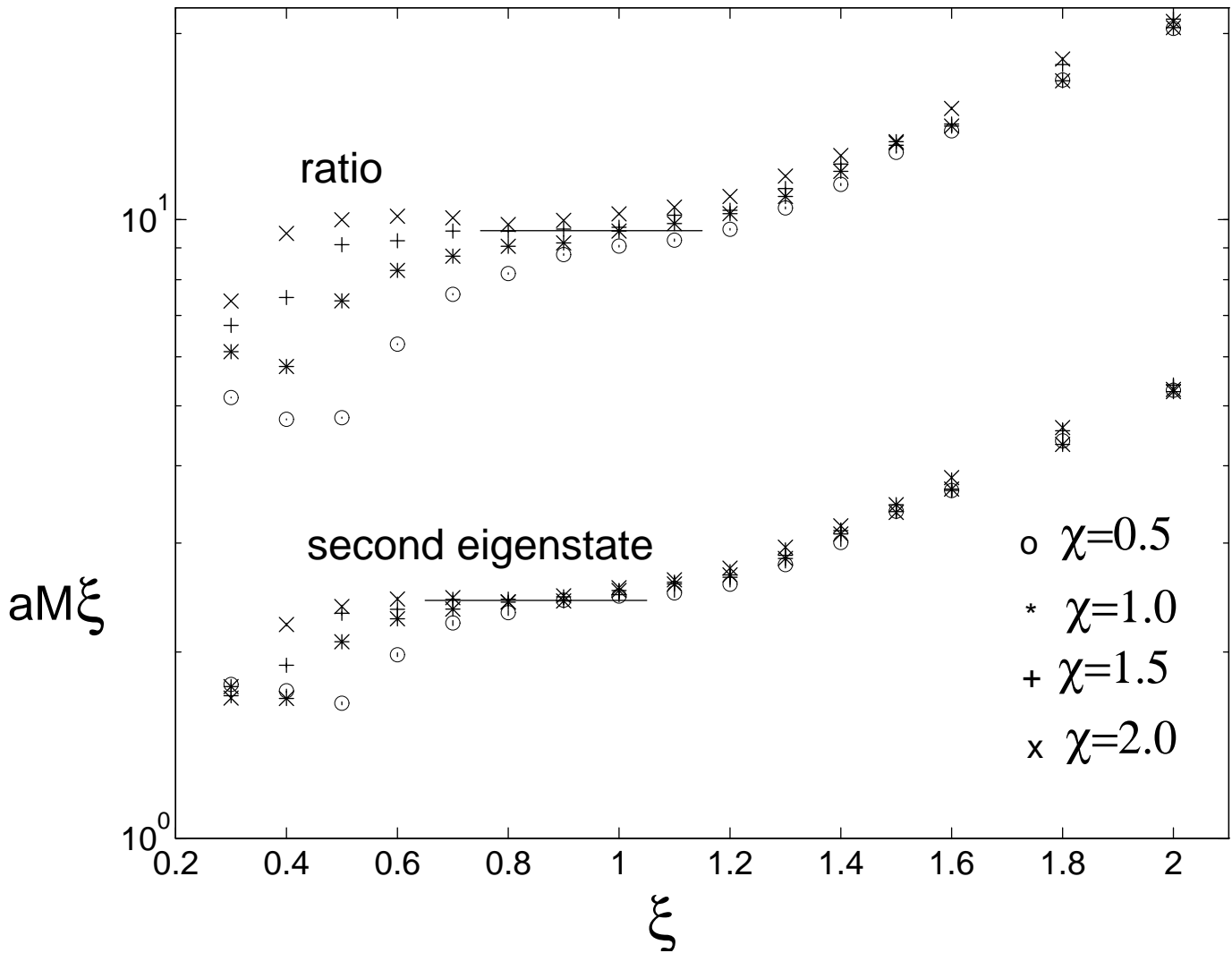

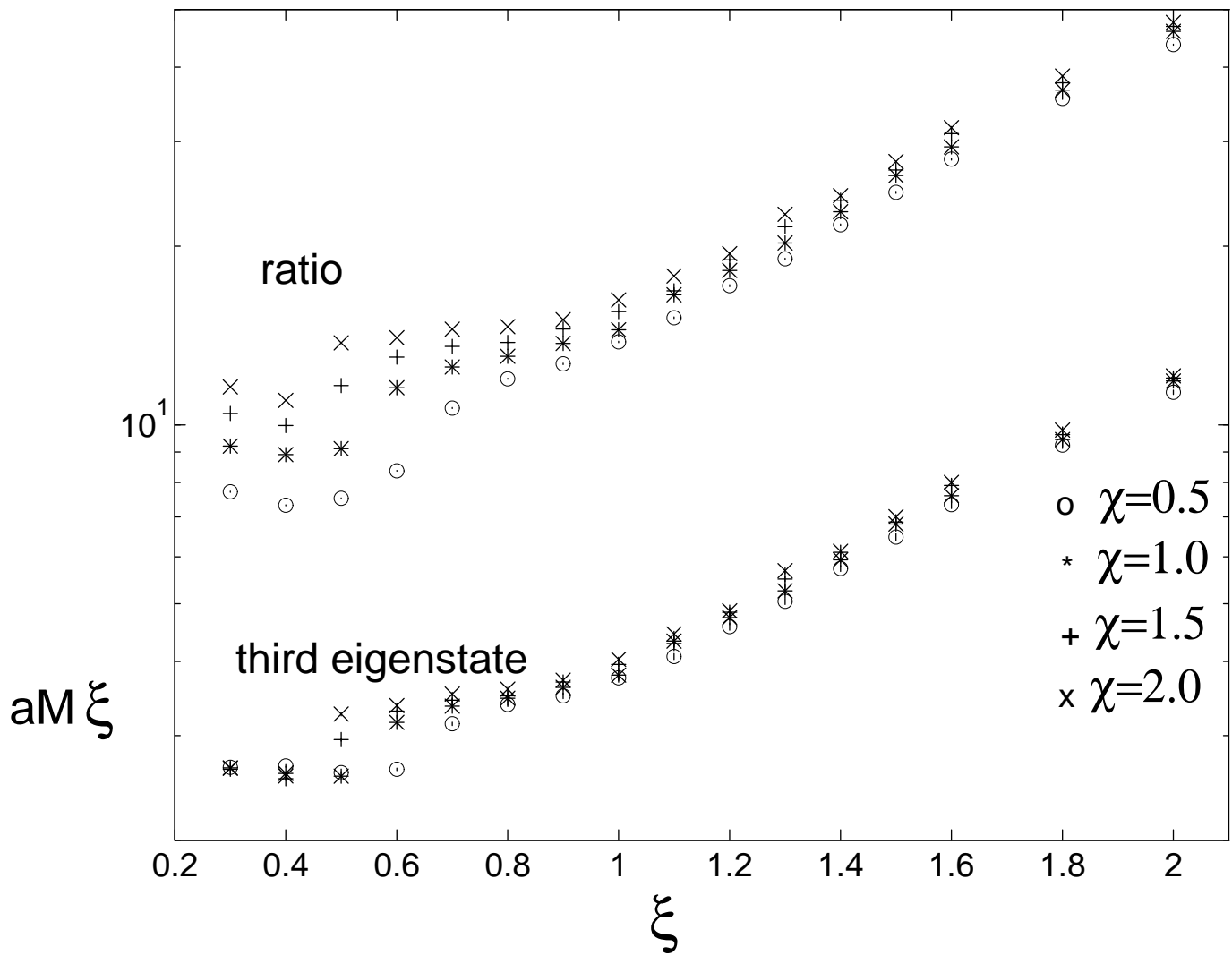

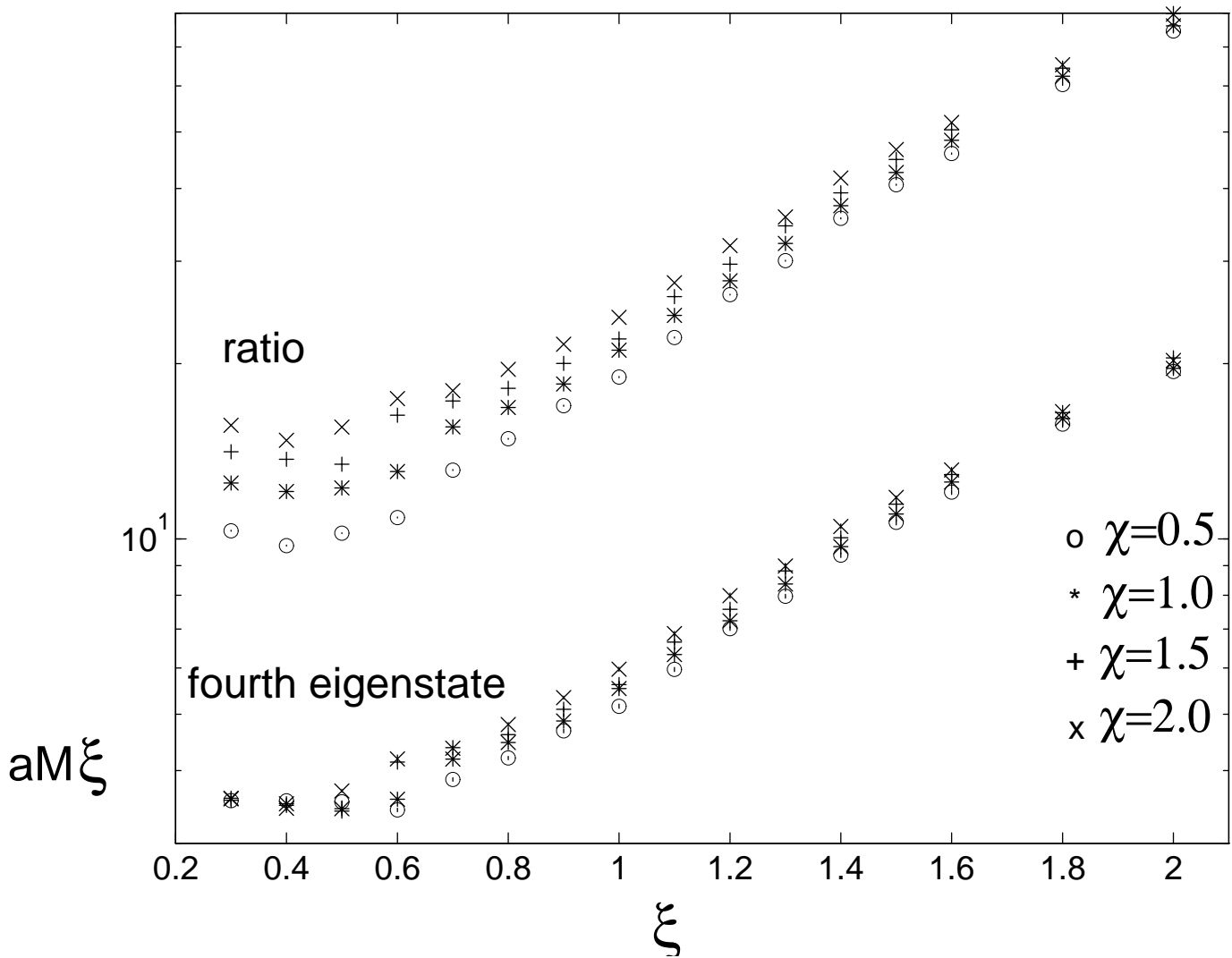